# Observation of the zero Hall plateau in a quantum anomalous Hall insulator

Yang Feng[1,*], Xiao Feng[1,*], Yunbo Ou[1,2], Jing Wang[3], Chang Liu[1], Liguo Zhang[1], Dongyang Zhao[1], Gaoyuan Jiang[1], Shou-Cheng Zhang[3,4], Ke He[1,4,†], Xucun Ma[1,4], Qi-Kun Xue[1,4], Yayu Wang[1,4,†]

[1]*State Key Laboratory of Low Dimensional Quantum Physics, Department of Physics, Tsinghua University, Beijing 100084, P. R. China*

[2]*Institute of Physics, Chinese Academy of Sciences, Beijing 100190, P. R. China*

[3]*Department of Physics, Stanford University, Stanford, CA 94305–4045, USA*

[4]*Collaborative Innovation Center of Quantum Matter, Beijing, China*

*\* These authors contributed equally to this work.*

[†] Emails: kehe@tsinghua.edu.cn; yayuwang@tsinghua.edu.cn

**Quantum anomalous Hall (QAH) effect in magnetic topological insulator (TI) is a novel transport phenomenon in which the Hall resistance reaches the quantum plateau in the absence of external magnetic field[1-3]. Recently, this exotic effect has been discovered experimentally in an ultrathin film of the $Bi_2Te_3$ family TI with spontaneous ferromagnetic (FM) order[3-6]. An important question concerning the QAH state is whether it is simply a zero-magnetic-field version of the quantum Hall (QH) effect, or if there is new physics beyond the conventional paradigm. Here we report experimental investigations on the quantum phase transition between the two opposite Hall plateaus of a QAH insulator caused by magnetization reversal. We observe a well-defined plateau with zero Hall conductivity over a range of magnetic field around coercivity, consistent with a recent theoretical prediction[7]. The features of the zero Hall plateau are shown to be closely related to that of the QAH effect, but its temperature evolution exhibits quantitative differences from the network model for conventional QH plateau transition[8]. We propose that the chiral edge states residing at the magnetic domain boundaries, which are unique to a QAH insulator, are responsible for the zero Hall plateau. The rich magnetic domain dynamics makes the QAH effect a distinctive class of quantum phenomenon that may find novel applications in spintronics.**

The realization of QAH effect requires that a two-dimensional (2D) material must be FM, topological, and insulating simultaneously[9]. Magnetically doped TIs have been proposed[1, 2, 10-12] and experimentally proved[3-6] to be an ideal material system for fulfilling these stringent requirements. For a 3D TI, the inverted bulk band structure ensures topologically protected metallic surface states (SSs), which become 2D when the film is sufficiently thin[13]. The spontaneous FM order induced by magnetic doping not only leads to the anomalous Hall effect, but also opens an energy gap at the Dirac point. When the Fermi level ($E_F$) lies within this gap, the only remaining conduction channel is the quasi-one-dimensional chiral edge state, which gives rise to quantized Hall resistance and vanishing longitudinal resistance at zero magnetic field[3, 14]. Up to date, the QAH effect has been observed in Cr or V doped $(Bi,Sb)_2Te_3$ TI thin films with accurately controlled chemical composition and thickness grown by molecular beam epitaxy (MBE)[3-6].

The MBE-grown QAH insulator film studied here has a chemical formula $Cr_{0.23}(Bi_{0.4}Sb_{0.6})_{1.77}Te_3$ and thickness $d$ = 5 QL (quintuple layer). Shown in Fig. 1a is a schematic drawing of the transport device, which is similar to that reported previously[3]. The film is manually scratched into a Hall bar geometry, and the $SrTiO_3$ substrate is used as the bottom gate oxide due to its large dielectric constant at low temperature. The Cr concentration, hence the density of local moment, is higher than that in the sample where the QAH effect was originally discovered[3]. As a result, the FM order forms at a higher Curie temperature $T_C$ = 24 K as determined by the temperature dependent anomalous Hall effect (supplementary Fig. S1). Another important consequence of higher Cr doping is that the sample becomes more disordered, which is crucial to the physics that will be discussed in this work.

We first demonstrate the existence of QAH effect in this sample. Fig. 1b displays the gate voltage ($V_g$) dependence of the Hall resistance $\rho_{yx}$ (blue curve) and longitudinal resistance $\rho_{xx}$ (red curve) measured at $T$ = 10 mK in a strong magnetic field $B$ = 12 T applied perpendicular to the film. The $\rho_{yx}$ exhibits a plateau for -10 V < $V_g$ < 10 V with its maximum value close to 99.1% of the quantum resistance $h/e^2$ ~ 25.8 kΩ. In the same

$V_g$ range $\rho_{xx}$ shows a pronounced dip with its minimum value close to 0.1 $h/e^2$. To show that the apparent Hall quantization in Fig. 1b is due to the QAH effect rather than conventional QH effect in high magnetic field, in Fig. 1c we display the field dependence of $\rho_{yx}$ measured at $V_g$ = -5 V, when $\rho_{xx}$ reaches a minimum in Fig. 1b. The Hall trace shows an abrupt jump at zero magnetic field, characteristic of the anomalous Hall effect. With increasing magnetic field, the $\rho_{yx}$ value increases gradually and approaches $h/e^2$ at 12 T. The $\rho_{xx}$ shown in Fig. 1d exhibits two sharp peaks at the coercive field $H_C$, and decreases rapidly on both sides. The $\rho_{xx}$ value is around 3 $h/e^2$ at zero field, but is significantly suppressed to 0.1 $h/e^2$ at $\mu_0 H$ = 12 T. The overall trend is the same as that shown in the original report of QAH effect[3], although the quantization is less complete.

To reveal the nature of the quantum phase transition between the two opposite QAH plateaus, in Figs. 2 we zoom into the low field part of the curves in Fig. 1c and 1d. Although the $\rho_{yx}$ curve (Fig. 2a) shows the typical square-shaped hysteresis, the zero field $\rho_{yx}$ value is merely 0.93 $h/e^2$ and the transition near $H_C$ is rather smooth. In the butterfly-shaped $\rho_{xx}$ curve (Fig. 2b), the peak $\rho_{xx}$ value at $H_C$ reaches 35 $h/e^2$, much larger than that reported previously[3]. Both observations can be explained by the high degree of disorder and small domain size in the current sample with larger Cr concentration, which is not ideal for the achievement of QAH effect. A highly intriguing phenomenon can be revealed when we convert the measured resistivity to longitudinal and Hall conductivity $\sigma_{xx}$ and $\sigma_{xy}$ by using the tensor relation:

$$\sigma_{xx} = \frac{\rho_{xx}}{\rho_{xx}^2 + \rho_{yx}^2} \text{ and } \sigma_{xy} = \frac{\rho_{yx}}{\rho_{xx}^2 + \rho_{yx}^2}.$$

Fig. 2c displays the $\sigma_{xx}$ and $\sigma_{xy}$ converted from the resistivity data shown in Fig. 2a and 2b. At $H_C = 0.16$ T, the longitudinal conductivity (red curve) shows a minimum with $\sigma_{xx}$ ~ 0.03 $e^2/h$, reflecting the highly insulating behavior when the magnetic domains reverse. The Hall conductivity (blue curve), interestingly, shows a plateau with $\sigma_{xy} = 0$ spanning a field scale of ~0.12 T centered around $H_C$. This zero Hall plateau, which was theoretically predicted in Ref. 7, has never been observed before by experiment and is the main focus of the current work.

To unveil the origin of the zero Hall plateau, in Fig. 3 we display the magnetic field dependent $\sigma_{xy}$ curves measured under varied gate voltages at $T = 50$ mK (the raw data of $\rho_{xx}$ and $\rho_{yx}$ are displayed in Fig. S2). For $V_g$ between -10 V and +10 V (Fig. 3b to 3f), there is always a well-defined zero Hall plateau around $H_C$. This gate voltage range corresponds to the QAH plateau in Fig. 1b, when $E_F$ lies within the magnetic gap at the Dirac point. When the $V_g$ value is increased to ±30 V (Fig. 3a and 3g), which are outside the QAH regime, the zero Hall plateau feature is much weakened and the Hall conductivity merely shows a shoulder-like curvature around $H_C$. This gate voltage dependence clearly indicates that the zero Hall plateau is closely related to the QAH effect.

The same conclusion can be drawn from the temperature evolution of the zero Hall plateau. Shown in Fig. 4a and 4b are the conductivity measured at seven different temperatures under a fixed $V_g = -5$ V (the raw data of $\rho_{xx}$ and $\rho_{yx}$ are displayed in Fig. S3). The $\sigma_{xx}$ curves in Fig. 4a show that with increasing temperature, the $H_C$ decreases and the minimum $\sigma_{xx}$ value at $H_C$ increases. This is due to the increase of thermally activated parallel SS conduction at higher temperatures. The $\sigma_{xy}$ curves in Fig. 4b show that the zero Hall plateau near $H_C$ narrows at high temperatures, and becomes barely visible at $T = 3$ K. This is accompanied by the decreases of high field $\sigma_{yx}$ value due to the weakening of QAH effect at high temperatures. In Fig. 4c we plot the temperature dependence of the zero Hall plateau width and the $\rho_{yx}$ value at 12 T extracted from Fig. S3. The two quantities closely track each other, again demonstrating the intimate correlation between the zero Hall plateau and the QAH effect.

Both the gate voltage and temperature dependence indicate that the zero Hall plateau originates from the same physical mechanism as the QAH effect, namely the quantum transport by the spontaneously generated chiral edge states. The difference is that in the QAH regime, all the magnetic domains in the FM TI are aligned along the same direction, thus there is only one pair of chiral edge states propagating along the sample boundary[11]. The zero Hall plateau, on the contrary, occurs around the coercive field when the magnetic domains reverse. Near $H_C$ there are a large number of randomly distributed upward and downward domains, leading to the proliferation of chiral edge states at the domain walls. The phenomenology of the zero Hall plateau should be explained by considering the transport properties of this network of chiral edge states.

Recently, a microscopic model has been proposed to describe the critical properties of the plateau transition in a QAH insulator[7]. The basic physical picture is that at the coercive field, the chiral edge states at the magnetic domain walls may tunnel into each other when the distance between them is shorter than the spatial decay length of the edge state. Therefore, the plateau transition at the coercivity in a QAH insulator can be mapped to the network model of the integer QH plateau transition in the lowest Landau level

(LL)[8, 15-17]. It was predicted that an intermediate plateau with zero Hall conductivity could occur at $H_C$, and the longitudinal conductivity will show two peaks[7]. The mechanism for the zero Hall plateau is that during magnetization reversal near $H_C$, the mean value of the exchange field gap induced by FM order approaches 0. Accordingly, the system is transitioned into an insulating state where the first Chern number becomes $C_1 = 0$ and $\sigma_{xy} = C_1 \cdot e^2/h = 0$.

Although the zero Hall plateau feature observed here is qualitatively consistent with the theoretical prediction in Ref. 7, a closer examination of the data reveals quantitative differences. First, the $\sigma_{xx}$ curves shown in Fig. 4a reach a minimum at $H_C$ and increase monotonically on both sides. This is inconsistent with the double-peak structure predicted in the theory[7]. More importantly, the temperature evolution of the $\sigma_{xy}$ curves shown in Fig. 4b cannot be described by the scaling law laid out in the theory[18, 19]. The blue open circles in Fig. 4d depict the temperature dependence of the slope of the $\sigma_{xy}$ vs. $H$ curve just outside the zero Hall plateau. It decreases gradually with lowing $T$, and becomes almost a constant below $T = 200$ mK. This is opposite to the theoretical prediction of the divergence of the slope as $T$ decreases towards zero (Ref. 7).

One reason for the inconsistency between theory and experiment is the sample studied here is not as perfect as considered theoretically due to the residual SS conduction. For a more ideal QAH insulator (Fig. S4a), the $\sigma_{xx}$ curve indeed shows a weak but visible double-peak structure (Fig. S4b upper panel). However, in such a sample the zero Hall plateau in $\sigma_{xy}$ is much weakened, only showing a slight curvature near $H_C$ (Fig. S4b lower panel). This is because the occurrence of zero Hall plateau requires the sample to break up into many small magnetic domains at $H_C$ to create a large number of chiral edge states at domain walls. The current sample with high Cr doping and strong disorder favors the domain proliferation, but an ideal QAH sample with weak disorder prefers the formation of a small number of large domains during magnetization reversal. It is thus fundamentally prohibitive to optimize the zero Hall plateau and QAH plateau simultaneously, although their physical mechanisms are closely related.

The temperature dependence of the slope of $\sigma_{xy}$ shown in Fig. 4d is more puzzling. In the network model for conventional QH effect, the plateau transition becomes steeper at lower temperature due to the increase of phase coherence length[20-22]. We propose that the most likely cause of the anomalous behavior here is due to the magnetic domain dynamics in a QAH insulator. For a FM material, the shape of the magnetic hysteresis loop is a very complicated issue that depends on various factors including magnetic anisotropy, strength of the exchange coupling and homogeneity of the sample[23]. In the current sample, the steeper slope of $\sigma_{xy}$ at higher temperature indicates that the reversal of magnetic domain becomes easier with increasing $T$, although its exact reason is unclear at the moment. The dominant factor here is thus not the phase coherence length of the edge states, but instead the magnetic domain dynamics of the FM TI. Such physics is certainly not included in the network model for the QH plateau transition, and is a unique feature of the QAH plateau transition that is yet to be treated rigorously by theory.

The zero Hall plateau state discovered here is absent in ordinary 2D electron systems with parabolic dispersion because no zeroth LL is allowed. Even for 2D Dirac fermion system with zeroth LL, the observed zero Hall plateau[24, 25] is microscopically different from the multi-domain network of chiral edge states discussed above. For example, the diverging $\rho_{xx}$ at the zero-energy state in graphene QH state is found to have a field-induced metal-insulator transition with Kosterlitz-Thouless type[26]. Whereas for the zero Hall plateau state here, the $T$ dependence of the peak $\rho_{xx}$ value at $H_C$ exhibits variable range hopping behaviour above 200 mK (see the fit in Fig. 4d). This is due to the suppression of tunneling between the chiral edge states at low temperature when the localization length is small enough. The deviation from the variable range hopping fit below 200 mK suggests that the sample enters a new insulating ground state whose nature is still unknown.

The observation of the zero Hall plateau at the coercivity of a QAH insulator shed new lights on the nature of the quantum phases and phase transitions in this exotic state of matter. From the quantum phase transition perspective, the QAH effect may not be a simple zero-field version of the QH effect due to the unique domain wall dynamics. A new theoretical model is required to describe its critical behaviour around the coercive field. From the application point of view, in recent years there are significant progresses in utilizing the nano-sized domains in FM materials for achieving new spintronic devices, such as the race track memory[27]. The quantum transport properties of the chiral edge states, in conjunction with the rich magnetic domain dynamics of a QAH insulator, may give it unique advantages for future spintronic applications.

**Methods**

**MBE sample growth**

The 5 QL $Cr_{0.23}(Bi_{0.4}Sb_{0.6})_{1.77}Te_3$ film is grown on an insulating $SrTiO_3(111)$ substrate by the same MBE growth method as described in our previous report[3]. Before sample growth, the substrate is degassed at 550 ℃ for 10 min and then heated at 650 ℃ for 25 min in the MBE chamber. High purity Bi (99.9999%), Sb (99.9999%), Cr (99.999%), and Te (99.9999%) are evaporated with commercial Knudsen cells. The growth was conducted under Te-rich conditions at a substrate temperature of 200 ℃ with a typical Te/(Bi,Sb) flux ratio of 10.

**Transport measurements**

The transport measurements are performed in a dry dilution refrigerator with a base temperature $T = 10$ mK and a magnetic field up to 12 T. The film is manually scratched into a Hall bar geometry, as described previously[3], and the electrodes are made by mechanically pressing small pieces of indium onto the contact areas of the film. The longitudinal and Hall resistances are measured by using standard four-probe ac lock-in method with an excitation current $I = 1$ nA.

**Acknowledgements**

This work was supported by the National Natural Science Foundation of China and the Ministry of Science and Technology of China. J. W. and S. C. Z. acknowledge the support from the U.S. Department of Energy, Office of Basic Energy Sciences, Division of Materials Sciences and Engineering, under contract No. DE-AC02-76SF00515, and from the FAME Center, one of six centers of STARnet, a Semiconductor Research Corporation program sponsored by MARCO and DARPA.

**Author contributions**

Y.Y.W., K.H. and Q.K.X. designed the research. X.F., Y.B.O., L.G.Z. and G.Y.J carried out the MBE sample growth. Y.F., C.L. and D.Y.Z. carried out the transport measurements. J.W. and S.C.Z. assisted in the theoretical calculations and X.C.M. assisted in the experiments. Y.F., C.L., J.W., K.H., and Y.Y.W. prepared the manuscript.

**Competing financial interests**

The authors declare no competing financial interests.

**Figure captions:**

**Figure 1 | Device schematic and the QAH effect. a**, Schematic drawing of the transport device used in this study. **b**, Gate voltage $V_g$ dependence of $\rho_{yx}$ (blue curve) and $\rho_{xx}$ (red curve) measured at $T = 10$ mK in a strong magnetic field $B = 12$ T applied perpendicular to the film. **c, d**, Magnetic field dependence of $\rho_{yx}$ (blue curve) (**c**) and $\rho_{xx}$ (red curve) (**d**) measured at $T = 50$ mK and $V_g = -5$ V.

**Figure 2 | Transition between the two opposite QAH plateaus at low magnetic field. a, b**, Low field part of the hysteretic anomalous Hall curve (blue curve) (**a**) and butterfly-shaped magnetoresistance (red curve) (**b**) measured at $T = 50$ mK and $V_g = -5$ V. **c**, Magnetic field dependence of $\sigma_{xx}$ (red curve) and $\sigma_{xy}$ (blue curve) converted from the resistivity data shown in (**a**) and (**b**), showing a minimum of $\sigma_{xx} \sim 0.03\ e^2/h$ and a Hall plateau with $\sigma_{xy} = 0$ around $H_C = 0.16$ T.

**Figure 3 | Magnetic field dependent $\sigma_{xy}$ curves measured under varied gate voltages at $T = 50$ mK.** (**a**) to (**g**): The $\sigma_{xy}$ for various $V_g$ values at -30 V, -10 V, -5 V, 0 V, +5 V, +10 V and +30 V, respectively. The zero Hall plateau clearly exists for $V_g$ between -10 V and +10 V and becomes weakened when $V_g$ value is increased to $\pm 30$ V, indicating that it is closely related to the QAH effect.

**Figure 4 | Temperature evolution of the zero Hall plateau. a, b**, The $\sigma_{xx}$ (**a**) and $\sigma_{xy}$ (**b**) measured at seven different temperatures under a fixed $V_g = -5$ V, showing that the $H_C$ decreases and zero Hall plateau narrows with increasing temperature. **c**, Temperature dependences of zero Hall plateau width and the $\rho_{yx}$ value at 12 T reveal intimate correlation between the zero Hall plateau and the QAH effect. **d**, Dependence of the logarithm of peak $\rho_{xx}$ value at $H_C$ (black open squares) and the maximum slope of the $\sigma_{xy}$ vs. $\mu_0 H$ curve near $H_C$ (blue open circles). The dashed red line shows the variable range hopping fit over one decade of temperature between 200 mK to and 3 K. Note both the $x$ and $y$ axes are in logarithmic scale.

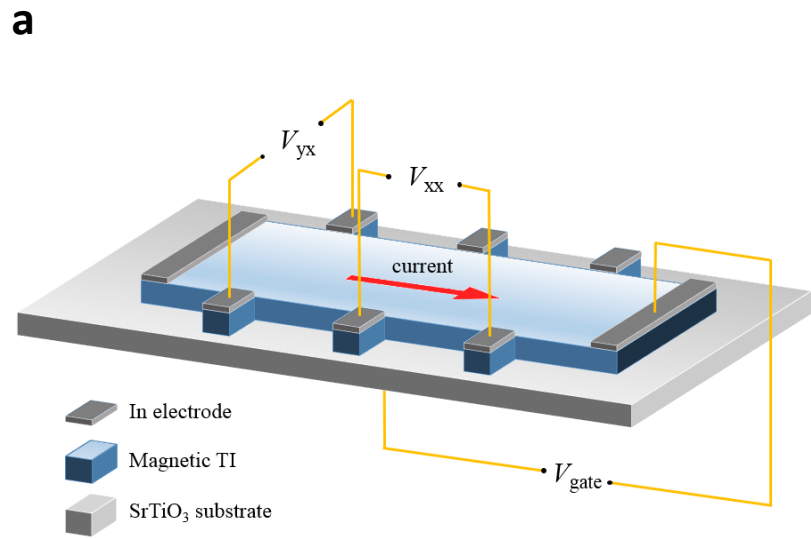
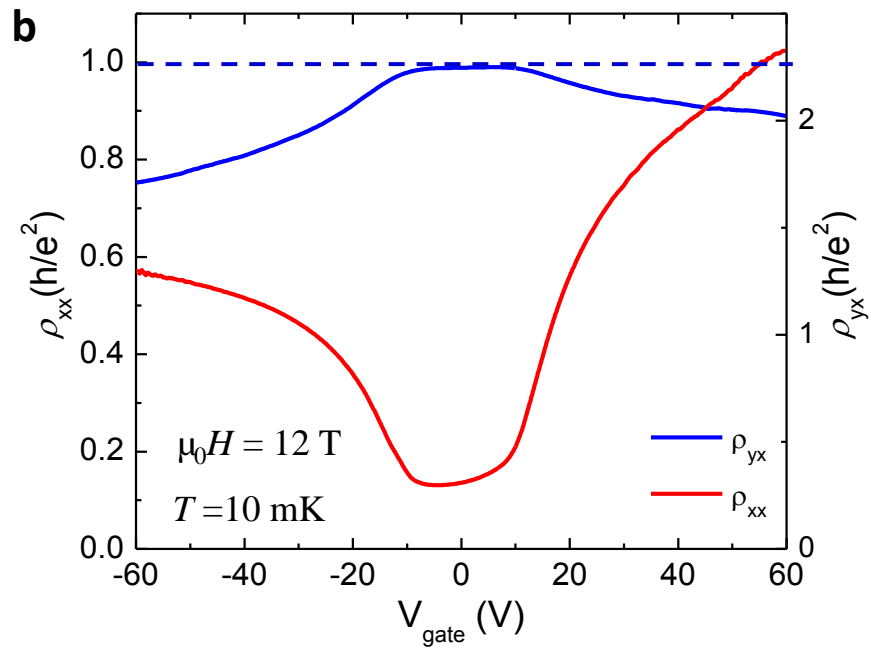
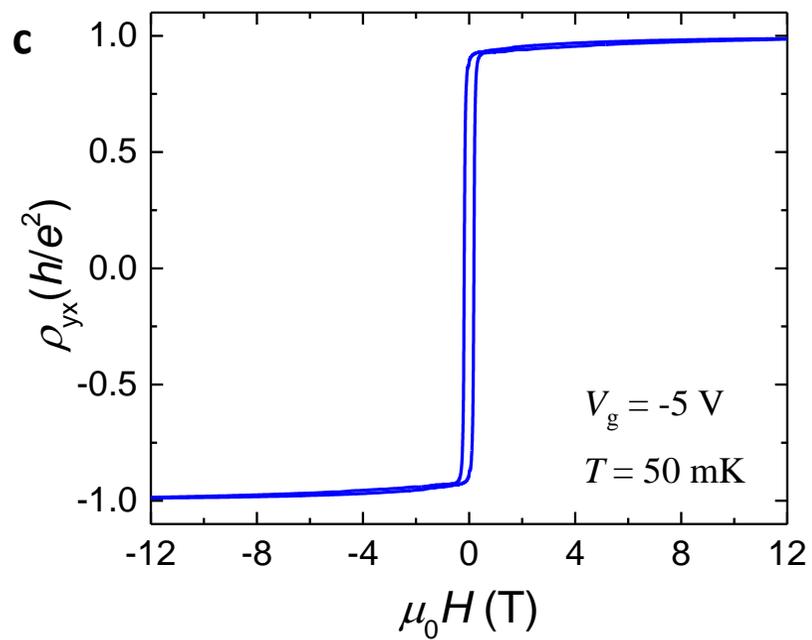
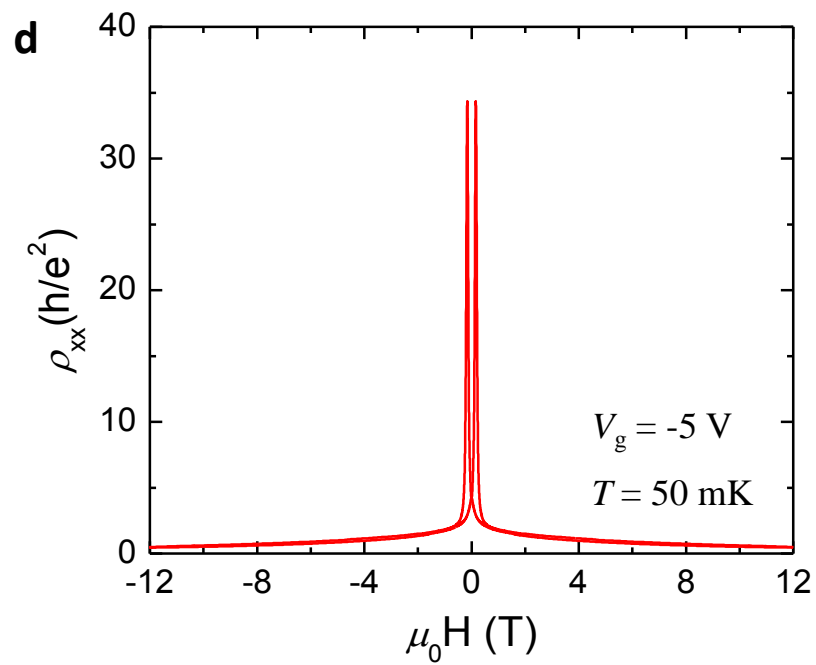

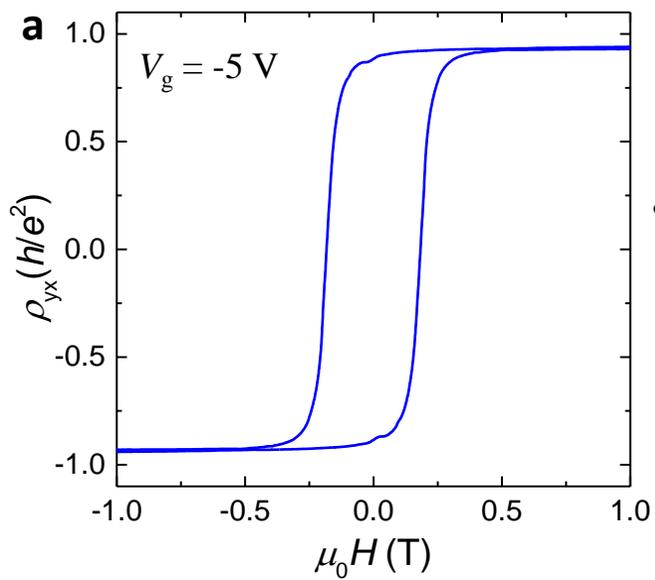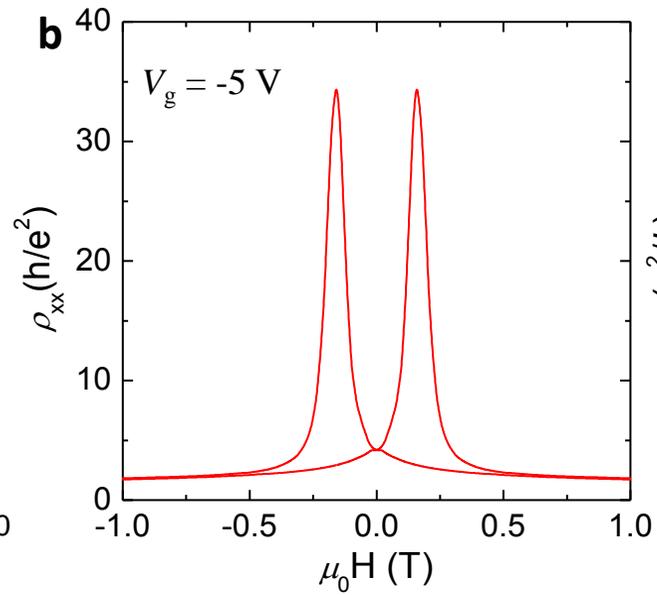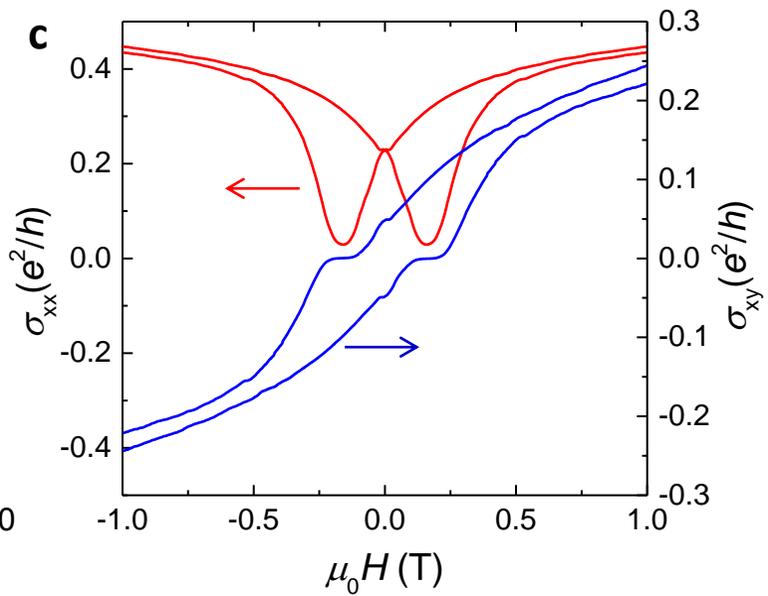

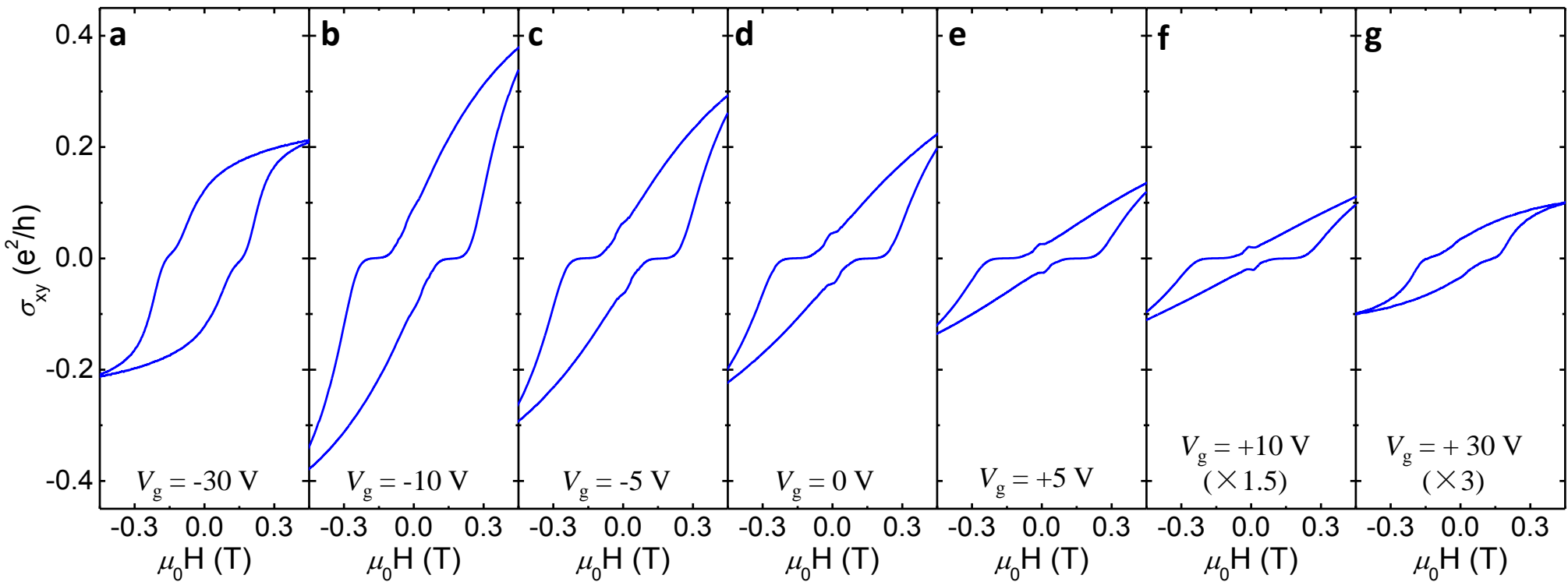

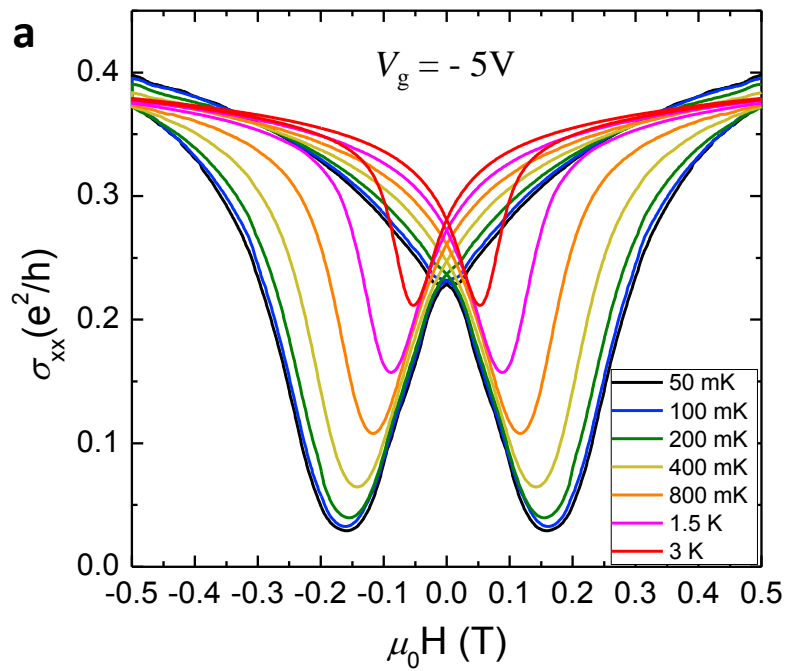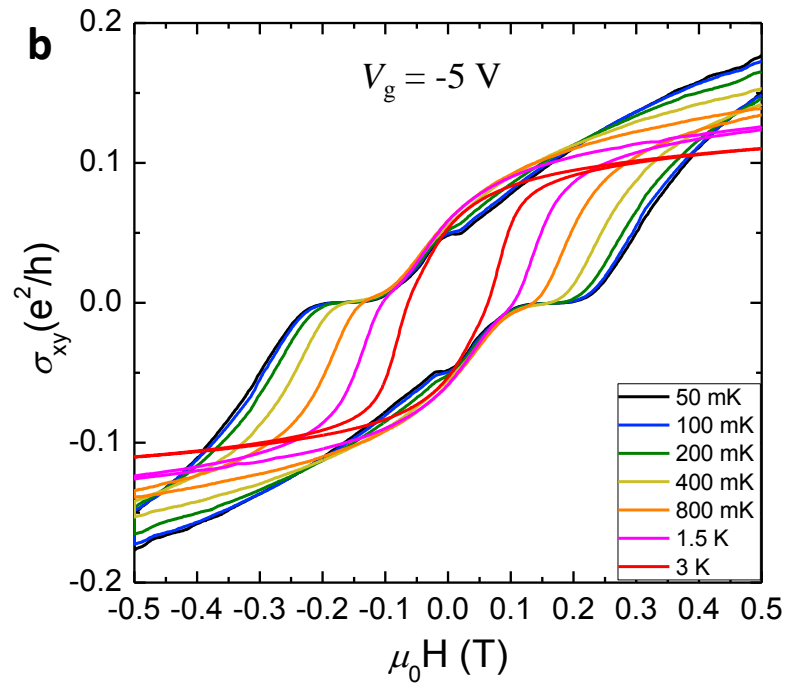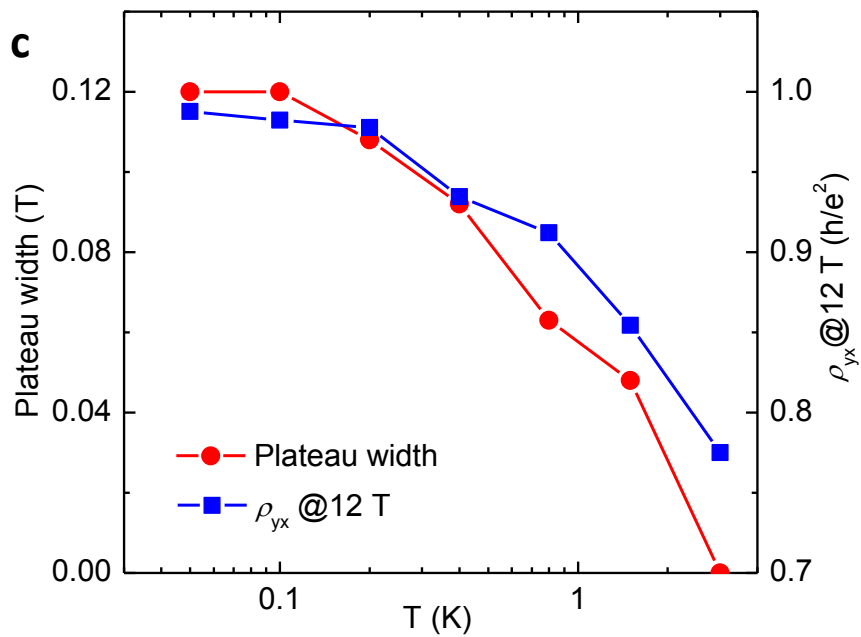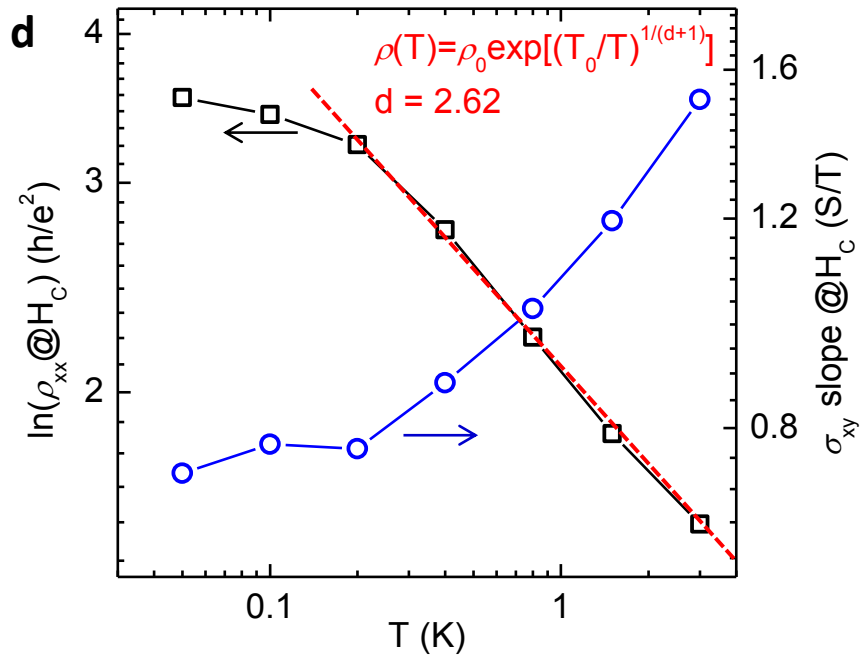

# Supplementary Information

# Observation of the zero Hall plateau in a quantum anomalous Hall insulator

Yang Feng[1,*], Xiao Feng[1,*], Yunbo Ou[1,2], Jing Wang[3], Chang Liu[1], Liguo Zhang[1], Dongyang Zhao[1], Gaoyuan Jiang[1], Shou-Cheng Zhang[3,4], Ke He[1,4,†], Xucun Ma[1,4], Qi-Kun Xue[1,4], Yayu Wang[1,4,†]

[1]*State Key Laboratory of Low Dimensional Quantum Physics, Department of Physics, Tsinghua University, Beijing 100084, P. R. China*

[2]*Institute of Physics, Chinese Academy of Sciences, Beijing 100190, P. R. China*

[3]*Department of Physics, Stanford University, Stanford, CA 94305–4045, USA*

[4]*Collaborative Innovation Center of Quantum Matter, Beijing, China*

*\* These authors contributed equally to this work.*

† Emails: kehe@tsinghua.edu.cn; yayuwang@tsinghua.edu.cn

**Contents:**

SI A: Determination of the Curie temperature

SI B: Raw Data of $\rho_{xx}$ and $\rho_{yx}$ for different gate voltages measured at $T$ = 50 mK

SI C: Raw data of $\rho_{xx}$ and $\rho_{yx}$ for various temperatures measured at $V_g$ = -5 V

SI D: The plateau transition in another QAH insulator with better quantization

**Figure S1 to S4**

**References**

## SI A : Determination of the Curie temperature

Figure S1a shows the magnetic field dependent Hall resistivity $\rho_{yx}$ of the 5 QL $Cr_{0.23}(Bi_{0.4}Sb_{0.6})_{1.77}Te_3$ film measured at different temperatures. The $\rho_{yx}$ in a ferromagnetic material is given by $\rho_{yx} = \rho_H B + \rho_{AH} (\mu_0 M)$, where the first term is the ordinary Hall resistivity and the second is the anomalous Hall contribution due to the magnetization $M$ of the material. The Curie temperature $T_C$ can be defined as the $T$ at which the anomalous Hall contribution decreases to zero. A common practice in determining the $T_C$ is to use the Arrott plot[1], in which $M^2$ is plotted against the ratio of $H/M$ ($H$ is the external magnetic field) at fixed $T$s. For each curve, the extrapolated intercept is proportional to the saturation magnetization, and the $T_C$ can be defined as when the intercept becomes zero. For the anomalous Hall effect, $\rho_{yx}$ is proportional to $M$, thus in the Arrott plot shown in Fig. S1b $M$ is replaced by $\rho_{yx}$. It is clear that at low $T$ the intercept is positive and becomes negative at high $T$. The intercept becomes close to zero at 24 K, which corresponds to the $T_C$ of this sample.

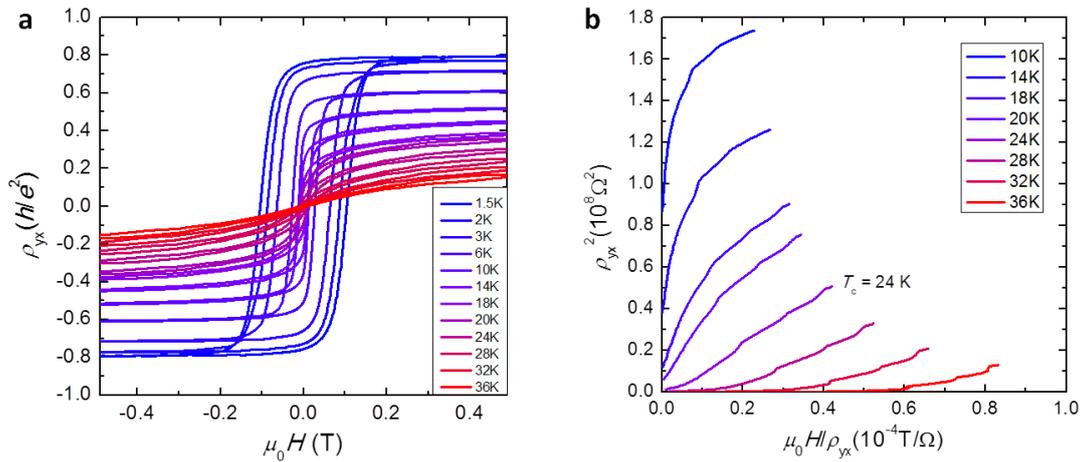

**Figure S1 | a**, The anomalous Hall effect curves taken at various temperatures from 1.5 K to 36 K. **b**, The corresponding Arrott plots of the $\rho_{yx}$ vs. $\mu_0 H$ isotherms in (**a**), showing the polarity change of the intercept with increasing temperatures, by which the Curie temperature can be extracted to be $T_C = 24$ K.

## SI B: Raw Data of $\rho_{xx}$ and $\rho_{yx}$ for different gate voltages measured at $T = 50$ mK

Figure S2 displays the Hall resistance $\rho_{yx}$ (upper panels) and longitudinal resistance $\rho_{xx}$ (lower panels) of the 5 QL $Cr_{0.23}(Bi_{0.4}Sb_{0.6})_{1.77}Te_3$ film measured at $T = 50$ mK under different bottom gate voltage $V_g$. The $\rho_{yx}$ curves all display the typical nearly square-shaped hysteresis, and the $\rho_{xx}$ curves show the butterfly shape with two peaks at the coercive field. In the $V_g$ range from -10 V to +10V, the anomalous Hall resistance is close to the quantum resistance. This corresponds to the regime that the Fermi level lies in the energy gap at the Dirac point and only the chiral edge states contribute to transport, leading to the QAH effect. Outside this $V_g$ range, the anomalous Hall resistance deviates quickly from the quantum resistance due to the parallel conduction from the surface states because now the Fermi level cuts through the surface band. The Hall conductivity curves shown in Fig. 3 of the main text are calculated from these curves.

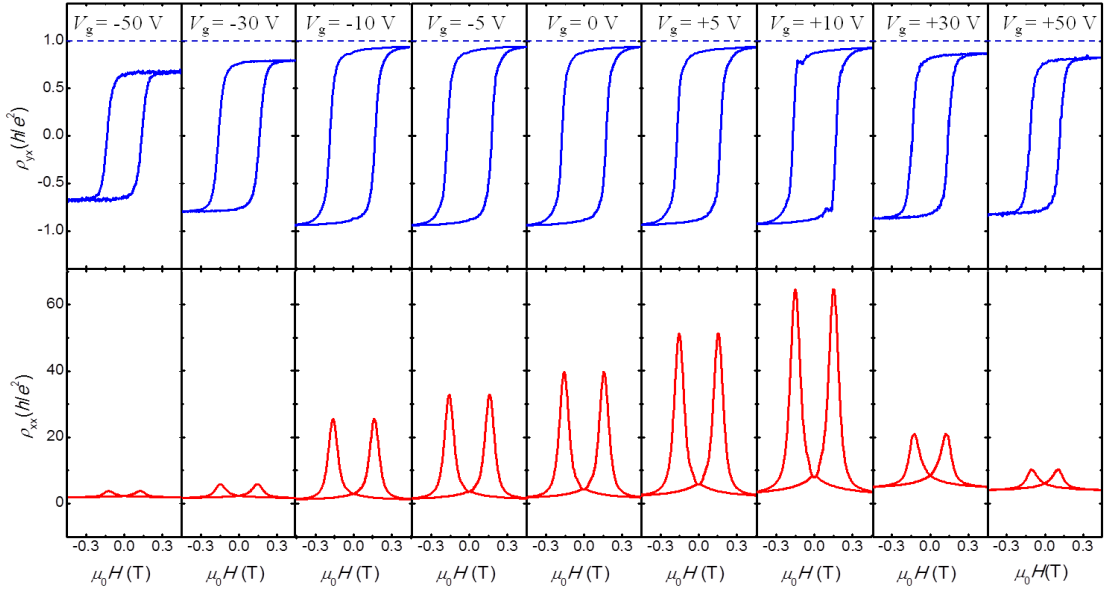

**Figure S2** | Magnetic field dependent Hall resistivity $\rho_{yx}$ (upper panel) and longitudinal resistivity $\rho_{xx}$ (lower panel) measured at $T = 50$ mK under varied gate voltages.

## SI C: Raw data of $\rho_{xx}$ and $\rho_{yx}$ for various temperatures measured at $V_g = -5$ V

Figure S3 displays a series of field dependent $\rho_{yx}$ and $\rho_{xx}$ curves measured at different temperatures far below $T_C$ under the same $V_g = -5$ V. Fig. S3a displays the low field part, clearly showing that the coercive field becomes larger with decreasing temperature. Meanwhile the peak $\rho_{xx}$ value at $H_C$ increases dramatically, as summarized in Fig. 4d of the main text. The conductivity curves shown in Fig. 4a and 4b in the main text are converted from these resistivity curves. Fig. S3b displays the high field data up to 12 T. With decreasing temperature, the high field $\rho_{yx}$ approaches the quantum resistance and $\rho_{xx}$ becomes closer to zero, indicating better Hall quantization. The $\rho_{yx}$ value at 12 T is extracted and plotted in Fig. 4c of the main text.

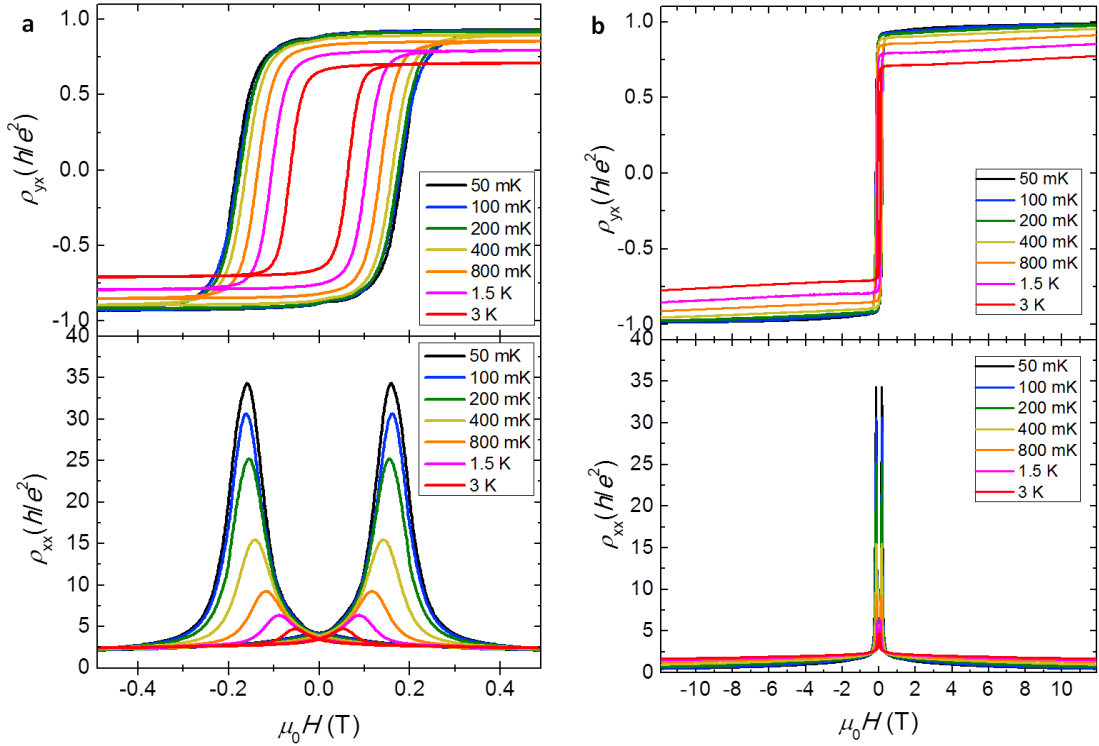

**Figure S3 | Temperature evolution of the measured resistivity curves.** The $\rho_{yx}$ and $\rho_{xx}$ curves under weak (**a**) and strong magnetic fields (**b**) measured at $V_g = -5$ V for $T = 50$ mK, 100 mK, 200 mK, 400 mK, 800 mK, 1.5 K and 3 K.

**SI D: The plateau transition in another QAH insulator with better quantization**

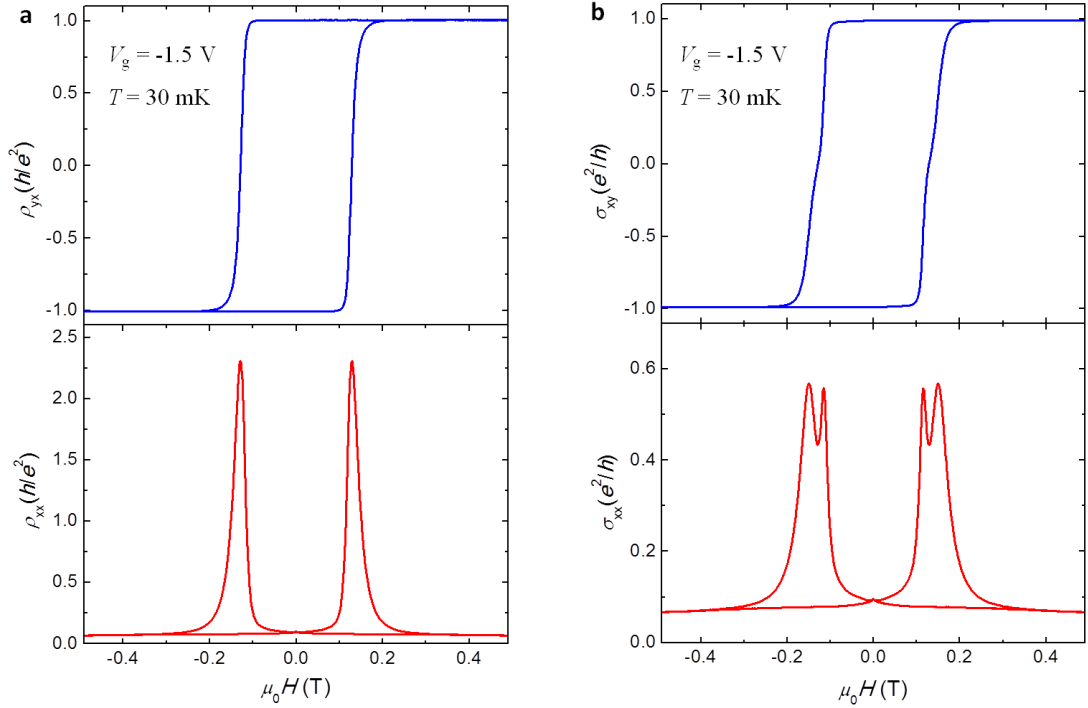

**Figure S4** | Magnetic field dependence of longitudinal conductivity $\sigma_{xx}$ (red curve) (**a**) and Hall conductivity $\sigma_{xy}$ (blue curve) (**b**), converted from the resistivity data of a more ideal QAH insulator.

In Fig. S4 we display the plateau transition in another QAH insulator with better Hall quantization. The sample is the one on which the QAH effect was originally observed[2], and has a chemical formula $Cr_{0.15}(Bi_{0.1}Sb_{0.9})_{1.85}Te_3$ so the Cr content is lower than the sample studied in this work. Fig. S4a shows the Hall and longitudinal resistivity, demonstrating a perfectly square-shaped anomalous Hall curve with the zero field $\rho_{yx}$ value already reaching the quantum resistance. The peak $\rho_{xx}$ value at $H_C$ is less than 10% of that in the current sample (see Fig. 2b in the main text), indicating much weaker disorder at coercivity presumably due to the larger magnetic domain size. Fig. S4b shows the converted conductivity curves. The $\sigma_{xx}$ curve exhibit a weak but visible double-peak structure at $H_C$, in agreement with the theoretical prediction[3]. However, the $\sigma_{xy}$ curve does not show the well-defined zero Hall plateau at $H_C$. Instead, it only exhibits a very weak wiggle. Therefore, the zero Hall plateau is less pronounced for the sample with more ideal QAH effect. As discussed in the main text, this is because the occurrence of zero Hall plateau requires the sample to break up into many small magnetic domains at $H_C$ to create a large number of chiral edge states, but an ideal QAH sample prefers the formation of a small number of large domains during magnetization reversal.